\documentclass[doublecol]{epl2} 
\usepackage{amssymb,bm}
\usepackage{amsmath,eucal}
\usepackage{graphics}

\title{Optimal stochastic transport in inhomogeneous thermal environments}
\shorttitle{Optimal stochastic transport in inhomogeneous thermal environments} %Insert here a short version of the title if it exceeds 70 characters

\author{Stefano Bo\inst{1,2,3,4} \and Erik Aurell\inst{2,5,6} \and Ralf Eichhorn\inst{3} \and Antonio Celani\inst{7}}
\shortauthor{S. Bo \etal}

\institute{

\inst{1}Cancer Cell Biophysics, IRC@C: Institute for Cancer Research at Candiolo, Str.~Prov.~142~km~3.95, 10060 Candiolo,
Torino, Italy\\    
\inst{2} Dept Computational Biology, AlbaNova University Centre, KTH - Royal Institute of Technology, SE-106 91 Stockholm, Sweden\\
\inst{3}Nordita, KTH Royal Institute of Technology and Stockholm University, Roslagstullsbacken 23, SE-106 91 Stockholm, Sweden\\
\inst{4} INFN, Sezione di Torino, via P. Giuria 1, 10125 Torino, Italy \\
\inst{5} ACCESS Linnaeus Centre, KTH - Royal Institute of Technology, SE-100 44 Stockholm, Sweden\\
\inst{6} Dept Information and Computer Science, Aalto University, PO Box 15400, FI-00076 Aalto, Finland \\
\inst{7} Physics of Biological Systems, Institut Pasteur and CNRS UMR 3525, 28 rue du docteur Roux, 75015 Paris, France}

\begin{document}

\abstract{
We consider optimization of the average
entropy production in inhomogeneous temperature environments
within the framework of stochastic thermodynamics. 
For systems modeled by Langevin equations (e.g. a colloidal particle in a heat bath)
it has been recently shown that a space dependent temperature breaks
the time reversal symmetry of the fast velocity degrees of freedom
resulting in an anomalous contribution to the entropy production of
the overdamped dynamics.
We show that optimization of entropy production is determined
by an auxiliary deterministic problem describing motion on a curved manifold in a potential. The ``anomalous contribution'' to entropy
plays the role of the potential
and the inverse of the diffusion tensor is the metric. 
We also find that entropy production is not minimized by adiabatically slow, quasi-static protocols
but there is a finite optimal duration for the transport process.
 As an example we discuss the case of a linearly space dependent diffusion coefficient.
}

\maketitle

\section{Introduction}
The last decades have witnessed a tremendous development in
our abilities to 
fabricate artificial devices on the micro- and nanometer scale,
and to manipulate and monitor biological and soft matter systems.
Out of the numerous evidences for this progress we mention
just two remarkable examples,
the realization of a micrometer-sized
Stirling-engine \cite{blickle11} and the verification of Landauer's
principle \cite{berut12} using a colloidal particle in a double-well
potential to represent the information memory.
Both these examples
link small non-equilibrium systems, in which
diffusive processes due to thermal fluctuations play a dominant role, 
to concepts well-known from macroscopic classical thermodynamics.
The theoretical basis for this connection is provided by
stochastic thermodynamics, a framework
which systematically extends thermodynamic quantities
such as exchanged heat, applied work \cite{sekimoto10} 
or entropy production
to individual fluctuating trajectories \cite{seifert12}.
For the distribution functions of such quantities,
exact general results can be
obtained,
the Jarzynski relation being probably the most prominent example
\cite{jarzynski97}.

For both the above mentioned experimental examples \cite{blickle11,berut12},
it is well-known that optimal bounds exist in the limit of adiabatically slow modulation
of the system: the Carnot efficiency for the Stirling engine \cite{blickle11},
and the Landauer bound for information erasure \cite{berut12}.
However, is it possible to find an optimal time-dependent ``control''
(realized by external forcings)
so that a specific quantity of interest becomes optimal during a process
which takes only finite time?
Within the framework of stochastic thermodynamics,
this question has first been posed by Seifert \cite{schmiedl07} in order to
minimize the mean work applied to a colloidal particle in a laser trap and to
calculate efficiency of finite-time working cycles \cite{schmiedl08}.
Afterwards it has been further extended to more general optimization problems
in a number of publications, see e.g.\
\cite{then08,geiger10,aurell11,aurell12,aurell12a,muratore12,sivak12}.

All these studies have been performed for systems in contact with a single heat
bath at constant temperature.
In many cases of interest, however, especially when considering Brownian
and molecular motors (see for example \cite{sekimoto10,prost97,reimann02}), transport is induced by systematically changing the 
temperature in time and/or by generating temperature gradients.
A recent work~\cite{zulkowski12} considered the case of a time varying (though spatially homogeneous) temperature 
which is used as an additional control parameter.
In the present Letter we study optimal finite-time processes in the presence of
temperature gradients by optimizing the total entropy
production \cite{aurell12a} of a system described by Langevin equations.
In doing so, we take into account that,
if temperature is not homogeneous in space, the correct
expression for the entropy production in the strong friction limit
is not simply given by the overdamped approximation of the entropy production
functional, but has an additional ``anomalous'' contribution stemming from a symmetry
breaking of the fast velocity degrees of freedom induced by the temperature gradient \cite{celani12}.

In an earlier work, it has been shown that for a constant diffusion matrix,
the control which optimizes heat or work is essentially given by the solution of an
auxiliary problem described by deterministic transport according to Burgers equation
\cite{aurell11}.
However, this is not the case any more if temperature is space dependent. 
We will show here that optimization of entropy production 
in inhomogeneous temperature environments
can still be mapped into a deterministic transport problem.
Furthermore, we find that for constant temperatures but space 
dependent friction coefficient the auxiliary problem is equivalent
to finding the geodesics on a curved manifold,
where the metric tensor is the inverse of the diffusion matrix.

\section{Entropy production in inhomogeneous media}
We consider driven diffusive motion in an inhomogeneous temperature environment 
modeled by the Langevin equation
\begin{eqnarray}
\label{eq:genModel}
\dot{\pmb{x}}(t) &=& \frac{\pmb{f}(\pmb{x},t)}{\gamma(\pmb{x})}+T(\pmb{x})\pmb{\nabla}\gamma^{-1}(\pmb{x})+
 \sqrt{\frac{2T(\pmb{x})}{\gamma(\pmb{x})}} \,\pmb{\eta}(t) 
\end{eqnarray}
where we have allowed temperature $T(\pmb{x})$ 
and friction coefficient $\gamma(\pmb{x})$ to be space-dependent
with stationary profiles.
The first term $\pmb{f}(\pmb{x},t)/\gamma(\pmb{x})$
on the right-hand side represents the external deterministic driving forces
acting on the particle, while the last term $\sqrt{2 T(\pmb{x})/\gamma(\pmb{x})}\,\pmb{\eta}(t)$
models the impact of thermal fluctuations by unbiased Gaussian white noise
with correlations $\langle \eta^i(t)\eta^j(s) \rangle = \delta^{ij}\delta(t-s)$ 
(we set Boltzmann's constant to unity).
This multiplicative noise term is interpreted in the non-anticipative It\^{o}-convention.
The unusual contribution $T\nabla\gamma^{-1}$
is a consequence of space-dependence of friction \cite{lau07} and results
from the small-inertia limit of the underlying Langevin-Kramers dynamics \cite{SIcelani12}.
The inhomogeneous heat bath is assumed to locally fulfill Einstein's relation
$D(\pmb{x}) =\mathbb{I} T(\pmb{x})/\gamma(\pmb{x})$
for the diffusion matrix $D(\pmb{x})$, which is proportional to the
identity matrix $\mathbb{I}$.

In the following, we briefly recapitulate known properties of the
entropy production associated with diffusive motion according to (\ref{eq:genModel}).
It can be shown that the entropy production in the environment is given by the sum
of two terms: a regular one and an anomalous one\cite{celani12}. The regular one 
is defined as the log-ratio of 
the probability $P$ of a specific forward path,
which is a solution $\pmb{x}(t)$ of (\ref{eq:genModel}), to the
probability $\tilde{P}$ for the occurrence of the 
backward path in the time-reversed overdamped dynamics \cite{chetrite07}.
The anomalous contribution accounts for the breaking of time-reversal symmetry in
velocities, induced by
the temperature gradient. It appears in the limit of vanishingly small inertia
of the full Langevin-Kramers dynamics, but would be overlooked in the naive
overdamped approximation when setting mass to zero \cite{celani12}.
The entropy production in the environment thus reads
\begin{eqnarray}
\label{eq:Senv}
S_{\mathrm{env}} &=&\log{\frac{P}{\tilde{P}}}+S_{\mathrm{anom}} =\\
&=&\int \frac{1}{T}(\pmb{f} - \pmb{\nabla} T) \circ \mathrm{d}\pmb{x}+S_{\mathrm{anom}}\, ,\nonumber
\end{eqnarray}
where the integral
is along the path $\pmb{x}(t)$ and the
product labeled by the open circle has to be evaluated
according to the midpoint rule (Stratonovich convention).
Note that in addition to the entropy production $\pmb{f}/T$ as an effect of
the external forces there is also a regular contribution
$-\pmb{\nabla} T/T$ from the spatial change of temperature along
the path.

The entropy of the system itself is defined
as \cite{seifert05}
\begin{equation}
\label{eq:Ssys}
S_{\mathrm{sys}} = -\ln \rho(\pmb{x},t) \, ,
\end{equation}
where $\rho(\pmb{x},t)$ is the solution of the Fokker-Planck equation
associated with (\ref{eq:genModel}). 
The total entropy production  (in the system and the environment)
along the path $\pmb{x}(t)$ is therefore
given by
\begin{eqnarray}
\label{eq:Stot}
S_{\mathrm{tot}} = \underbrace{\int \left [ -\mathrm{d}\ln \rho + \frac{1}{T}(\pmb{f} - \pmb{\nabla}T) \circ \mathrm{d}\pmb{x}  \right ]}_{ S_{\mathrm{reg}}}
+S_{\mathrm{anom}}\, .
\end{eqnarray}
Averaging (\ref{eq:Stot}) over many realizations of the path $\pmb{x}(t)$ with
given distribution $\rho_0(\pmb{x}_0)=\rho(\pmb{x}_0,t=0)$ of the initial points
$\pmb{x}_0=\pmb{x}(t=0)$, we find the quantity of main interest, the total average entropy production $\langle S_{\mathrm{tot}} \rangle$. 
It can be written as 
\begin{equation}
\label{eq:avStot}
\langle S_{\mathrm{tot}} \rangle = \int_0^\tau \mathrm{d}t \left\langle
\pmb{v}D^{-1}\pmb{v} + \frac{(n+2)}{6} \frac{(\pmb{\nabla}T)^2}{\gamma T} \right\rangle
\, ,
\end{equation}
where $n$ is the dimensionality of $\pmb{x}$ and
$\tau$ denotes the time at which the path $\pmb{x}(t)$ ends.
The first term represents the regular entropy production,
the second term is the anomalous contribution \cite{celani12}.
To obtain the specific form of the regular part from the expression in
(\ref{eq:Stot}), we have made use of the
Fokker-Planck equation for $\rho(\pmb{x},t)$ associated with (\ref{eq:genModel}),
which can be written in form of the transport equation
\begin{equation}
\label{eq:genFP}
\frac{\partial \rho}{\partial t} + \pmb{\nabla}(\pmb{v} \rho) = 0
\end{equation}
with the current velocity \cite{nelson67,guerra83}
\begin{equation}
\label{eq:genv}
\pmb{v} = \frac{1}{\gamma}\left(\pmb{f} - \pmb{\nabla} T-T\pmb{\nabla} \ln \rho\right) \, ,
\end{equation}
and using the partial integration ``trick''
$\langle \pmb{\nabla}h \rangle = - \langle h \, \pmb{\nabla} \ln\rho \rangle$,
which is valid for arbitrary functions $h=h(\pmb{x},t)$ (bounded at infinity).
Note that if temperature is constant (the regular case with $\pmb{\nabla} T=0$),
the average total entropy production \eqref{eq:avStot} is simply a quadratic
form of the current velocity \cite{chetrite07}.

From expression
(\ref{eq:avStot}) it is clear 
that the average entropy production is largely determined
by the evolution of the distribution of paths $\pmb{x}(t)$.
Extremal entropy production therefore
requires a specific ``optimal'' evolution of paths.
Such ``optimal'' evolution can be imposed on the system
by applying a suitable time-dependent protocol \cite{schmiedl07}
to control (parts of) the external potentials and forces.
In our model (\ref{eq:genModel}), the external control is
incorporated into the term $\pmb{f}(\pmb{x},t)$
by its explicit time-dependence.

In the remainder of the paper, we study the
conditions under which the total average entropy production
(\ref{eq:avStot}) becomes extremal.
This optimization problem is typically subject to constraints,
in particular the distribution $\rho_0(\pmb{x}_0)$ of initial points
is usually prescribed. Other additional constraints may be present as well,
like a specific final distribution $\rho(\pmb{x},t=\tau)$
or a specific value of the control at final time $\tau$.

\section{Optimization of entropy production}
We now study the problem of optimizing the average
entropy production $\langle S_{\mathrm{tot}} \rangle$, 
which is of the general form
\begin{equation}
\label{eq:genS}
\langle S_{\mathrm{tot}} \rangle
	= \left\langle \int \mathrm{d}t \, L(\pmb{x}(t),\pmb{v}(\pmb{x}(t),t),t) \right\rangle \, ,
\end{equation}
where $L$ contains the regular part of the entropy production and the anomalous one,
\begin{equation}
\label{eq:genL}
L = (D^{-1})_{ij}(\pmb{x}) \, v^i v^j - U(\pmb{x}) \, .
\end{equation}
The ``potential'' term is given by the anomalous entropy production
\begin{equation}
\label{eq:U}
U = - \frac{(n+2)}{6T\gamma} \frac{\partial T}{\partial x^i}\frac{\partial T}{\partial x^i} \, .
\end{equation}
The first step in our analysis consists in rewriting
the average in (\ref{eq:genS}) over many realizations
of the path along which the integral is performed into an
equivalent average over the distribution $\rho(\pmb{x},t)$, obeying \eqref{eq:genFP},
\begin{eqnarray}\label{eq:S0}
\langle S_{\mathrm{tot}} \rangle
& = & \int \mathrm{d}\pmb{x} \int \mathrm{d}t \, L(\pmb{x}(t),\pmb{v}(\pmb{x}(t),t),t) \rho(\pmb{x},t) \\
& = & \int \mathrm{d}\pmb{x}_0 \, \rho_0(\pmb{x}_0) \int \mathrm{d}t \, L(\xi(t;\pmb{x}_0),\dot{\xi}(t;\pmb{x}_0),t)\nonumber \, .
\end{eqnarray}
In the second step we have used the formal solution
\begin{equation}
\label{eq:rho}
\rho(\pmb{x},t) = \int \mathrm{d}\pmb{x}_0 \, \delta(\pmb{x}-\xi(t;\pmb{x}_0)) \rho_0(\pmb{x}_0)
\end{equation}
of the Fokker-Planck equation (\ref{eq:genFP}),
where $\xi(t;\pmb{x}_0)$ solves the auxiliary deterministic dynamics
\begin{equation}
\label{eq:aux}
\dot{\xi}(t;\pmb{x}_0) = \pmb{v}(\xi(t;\pmb{x}_0),t)
\end{equation}
and $\rho_0(\pmb{x}_0)$ is the distribution of the path starting points $\pmb{x}_0$.
As already mentioned, $\rho_0$ is typically specified by the problem at hand.
The optimal average entropy production is thus obtained by
extremizing the time-integral
\begin{equation}
\label{eq:Sxi}
S_\xi = \int_0^\tau \mathrm{d}t \, L(\xi(t;\pmb{x}_0),\dot{\xi}(t;\pmb{x}_0),t) \, 
\end{equation}
in the second line of (\ref{eq:S0}) for any given initial point $\pmb{x}_0$.
This corresponds to a standard variational problem
for the auxiliary trajectories $\xi(t;\pmb{x}_0)$,
with the ``trajectory-wise'' entropy production (\ref{eq:Sxi})
being identical to the cost function \cite{aurell12a}.
The integrand $L(\xi,\dot{\xi},t)$, as specified in (\ref{eq:genL}),
can be interpreted as the ``Lagrange-function'' of the problem.
It contains a kinetic-like term $(D^{-1})_{ij} \dot{\xi}^i \dot{\xi}^j$
in curved space and a potential-like term $U$.
The metric of the space is equivalent to the inverse of the diffusion
coefficient,
\begin{equation}
\label{eq:g}
g_{ij} = (D^{-1})_{ij} \, .
\end{equation}
The optimal solutions $\xi(t;\pmb{x}_0)$ of this variational problem 
solve the Euler-Lagrange equations
\begin{equation}
\label{eq:EL}
\ddot{\xi}^i + \Gamma^{i}_{jk} \dot{\xi}^j \dot{\xi}^k + \frac{1}{2}D^{ij}\frac{\partial U}{\partial \xi^j} = 0
\end{equation}
with the Christoffel-Symbols $\Gamma^{i}_{jk}$, defined as
$\Gamma^{i}_{jk}= \frac{1}{2} g^{im} \left( \frac{\partial g_{mk}}{\partial x^j} + \frac{\partial g_{mj}}{\partial x^k} - \frac{\partial g_{jk}}{\partial x^m} \right)$.

The relation (\ref{eq:EL}) and its implications for optimal
stochastic transport are the main results of this paper.
We remark that if the temperature profile is homogeneous the anomalous contribution vanishes ($U=0$)
and minimization of the entropy production is equivalent to finding
the geodesics for free deterministic motion in a space
with metric tensor (\ref{eq:g}).

According to (\ref{eq:rho}) and (\ref{eq:aux}) the optimal density $\rho(\pmb{x},t)$ is
transported along these auxiliary trajectories with a local velocity which
corresponds to the current velocity (\ref{eq:genv}) at this point.
It is remarkable that the auxiliary deterministic dynamics (\ref{eq:EL})
 on the curved manifold can be fully described in terms of the current velocity.
Once the auxiliary problem is solved, the external protocol acting on $\pmb{f}(\pmb{x},t)$
has to be adapted to generate the optimal transport according to
the solution of (\ref{eq:EL}).

Since the Euler-Lagrange equation (\ref{eq:EL}) is a second-order
differential equation, we need---in addition to the initial point $\pmb{x}_0$---%
a second condition to obtain a unique solution, e.g., we need to fix the initial
velocity or an intermediate point along the trajectory. This freedom
can be used to meet the additional constraints of the optimization problem.
For instance, to reproduce a given final density $\rho(\pmb{x},\tau)$ we
may choose the final points $\xi(\tau;\pmb{x}_0)$, so that the relation
$\rho(\pmb{x},\tau) = \int \mathrm{d}\pmb{x}_0 \, \delta(\pmb{x}-\xi(\tau;\pmb{x}_0)) \rho_0(\pmb{x}_0)$
is fulfilled. In that case, for a constant temperature (no anomalous potential term),
 the optimization problem is equivalent
to an optimal assignment problem mapping the initial density $\rho_0(\pmb{x}_0)$
to the final one $\rho(\pmb{x},\tau)$ with the quadratic cost function
$\int \mathrm{d}\pmb{x}_0 \, \rho_0(\pmb{x}_0) \int \mathrm{d}t \, (D^{-1})_{ij} \dot{\xi}^i \dot{\xi}^j$
\cite{aurell12a}. 
If, however, no additional constraints are specified we can exploit
the freedom of choosing a second condition for the solution of (\ref{eq:EL}) to perform
a further optimization step over the final densities $\rho(\pmb{x},\tau)$.

\section{Example: one-dimensional motion for a linear diffusion coefficient}
In order to highlight the influence of an inhomogeneous temperature it is
instructive to study the one-dimensional transport of a Brownian particle between given
initial and final states (as introduced by Seifert and Schmiedl in \cite{schmiedl08}).
We will here compare two cases: the anomalous one of a linear temperature profile and
a regular one where temperature is constant but friction is space dependent.
We know that the latter situation is solved by geodesics.
The difference between the two cases gets more marked 
as the transport time increases and the process is closer to quasi-static operation.

In one dimension, the auxiliary equation of motion (\ref{eq:EL})
reads
\begin{equation}
\label{eq:EL1d}
\ddot{\xi} - \frac{1}{2 D}\frac{dD}{d\xi}\dot{\xi_t}\dot{\xi_t} + \frac{1}{2}D\frac{d U}{d\xi} = 0 \; 
\end{equation}
where we remark that in the regular case of constant
 temperature the potential term vanishes ($U=0$). 
 \begin{figure}
 \begin{minipage}[b]{8cm}
  \centering
   \includegraphics[width=8.0cm]{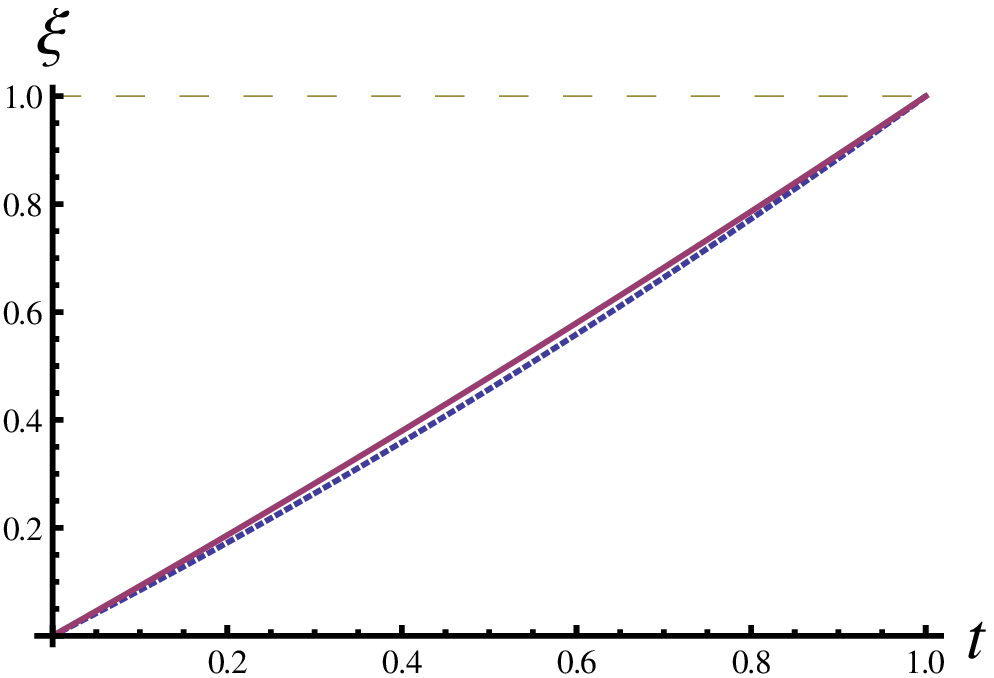}
 \end{minipage}

 \begin{minipage}[b]{8cm}
  \centering
   \includegraphics[width=8.0cm]{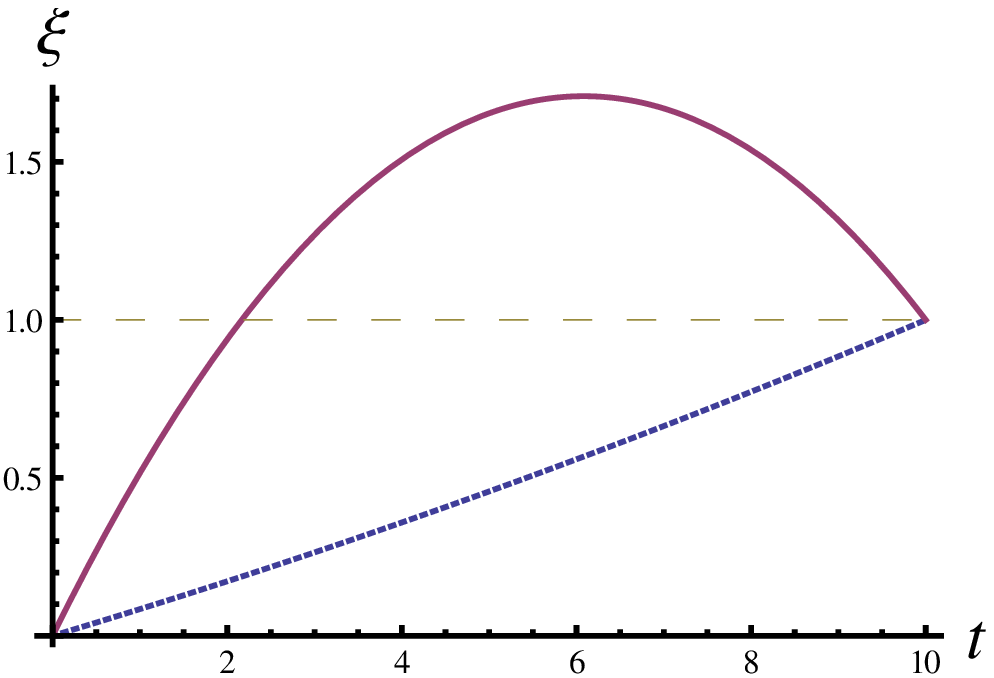}
 
 \end{minipage}
\caption{Optimal trajectory between position $x_0=0$ and and $x_\tau=1$ as a function of time
for $T_0=1$, $\vartheta=1$, $\gamma_0=1$.
The dotted line refers to the case without the entropic anomaly
and the solid line to the anomalous case of inhomogeneous temperature,
while the reference dashed line indicates the arrival position $x_\tau$.
Top: transport in a time $\tau=1$. Bottom: transport in a time $\tau=10$.}
\label{fig:compare}
\end{figure}
Since (\ref{eq:EL1d}) (like its general counterpart (\ref{eq:EL})) is obtained
from extremizing a ``Lagrange-function'' (\ref{eq:genL}), we actually
face a Hamiltonian dynamics with preserved ``energy'' $(D^{-1})_{ij} \dot{\xi}^i \dot{\xi}^j+ U$.
Therefore, (\ref{eq:EL1d})
can be easily solved for the quadratic velocity with the result
\begin{eqnarray} 
\dot{\xi}\dot{\xi} & = & \left( C- U(\xi) \right)D(\xi)
\nonumber \\
 & = & C D(\xi) + \frac{1}{2\gamma^2(\xi)} \left(\frac{d T(\xi)}{d \xi} \right)^2 \, ,
\label{eq:xidotxidot}
\end{eqnarray}
where $C$ is an integration constant corresponding to the conserved ``energy''.
For the thermally inhomogeneous study case we choose a linear temperature profile
and a constant friction coefficient
\begin{equation}
\label{eq:T}
T(x)=T_0+\vartheta x \, , \qquad \gamma=\gamma_0 \, ,
\end{equation}
with constant temperature gradient $\vartheta$. In order
to avoid non-physical zero or negative temperatures we
restrict to positions larger than $x_{\mathrm{min}}=-T_0/\vartheta$.
For the case of
space-dependent friction we can define
\begin{equation}
\label{eq:gamma}
T=T_0 \, , \qquad \gamma(x)=\frac{\gamma_0}{1+\vartheta/T_0\, x} \, .
\end{equation}
Therefore, in both cases, we have the same space dependent diffusion coefficient 
\begin{equation}
\label{eq:diff}
D(x)=\frac{T_0+\vartheta x} {\gamma_0} \, ,
\end{equation}
but only for the space-dependent temperature profile the second
term in (\ref{eq:xidotxidot}) contributes, representing the anomalous
entropy production.
With these definitions, the right-hand side of (\ref{eq:xidotxidot}) is linear in $\xi$
and can be solved explicitly
with a solution that is quadratic in time:
\begin{equation}
\label{eq:xi}
\xi(t;x_0) = x_0 - \bigl(x_0 + X\bigr) \frac{t}{\tau} + \bigl(x_\tau + X\bigr) \left( \frac{t}{\tau} \right)^2
\end{equation}
with
\begin{eqnarray}
\label{eq:X}
X = x_0 + \frac{2T_0}{\vartheta}
	- \sqrt{4\left(x_0+\frac{T_0}{\vartheta}\right)\left(x_\tau+\frac{T_0}{\vartheta}\right) + \chi \frac{\vartheta^2 \tau^2}{2\gamma_0^2}} \, ,
\end{eqnarray}
and the prescribed initial position $\xi(0;x_0)=x_0$
and final position $\xi(\tau;x_0)=x_\tau$, where for the sake of simplicity
we have considered $x_\tau>x_0$ (i.e. motion towards the hotter region)%
\footnote{By (\ref{eq:X}) the motion starts with a positive velocity.
There is a second solution with a plus sign replacing the minus sign in front of the square root
so that motion starts with negative velocities. It is easy to verify that these solutions
\emph{maximize} the average entropy production. However, this maximizing
trajectory, if unconstrained,
visits unphysical regions of positions corresponding to negative temperatures.}.
In (\ref{eq:X}),
we have introduced the characteristic parameter $\chi$ to
distinguish between the anomalous case with inhomogeneous
temperature (setup (\ref{eq:T}), $\chi=1$) and the regular
one with constant temperature but space-dependent
friction (setup (\ref{eq:gamma}), $\chi=0$).
Actually, the term involving $\chi$ in the square root
represents $-\tau^2 D U$ coming from \eqref{eq:xidotxidot}.

Beside its dependence on the parameters that define the specific
system (like $T_0$, $\vartheta$, and $\gamma_0$), and on the initial and final
positions $x_0$, $x_\tau$, this optimal solution also depends on the process
duration $\tau$. In the absence of the anomalous contribution to the entropy
production ($\chi=0$), this dependence corresponds to a trivial rescaling of
time. The optimal trajectory then is a parabola with positive curvature 
$x_\tau+X|_{\chi=0}$, independent of process duration.
In presence of the anomaly ($\chi=1$), however, the evolution of
$\xi(t;x_0)$ depends on $\tau$ also via $X$ and can change qualitatively.
For processes that take exactly $\tau=\sqrt{2}(x_\tau-x_0)\gamma_0/\vartheta$, 
the solution $\xi(t;x_0)$ is a straight line, while it is a parabola with
a positive (negative) concavity for shorter (longer) process durations.
Interestingly, for very slow processes with 
\begin{equation}\label{eq:overshhot}
 \tau>\tau_{\mathrm{o.s.}}=\frac{2\gamma_0}{\vartheta}\sqrt{2(x_\tau-x_0)(x_\tau+\frac{T_0}{\vartheta})}
\end{equation}
the optimal trajectory even overshoots the
target position at $x_\tau$ and eventually changes direction to finally reach it.
Such a counter-intuitive behavior can be traced back to the influence
of the anomalous contribution to the ``cost'' of the optimal trajectory.

In the present case, the entropic cost of the optimal evolution
reads (see eqs.~(\ref{eq:genL}), \eqref{eq:U}, (\ref{eq:Sxi}) and \eqref{eq:xidotxidot})
\begin{eqnarray}
\label{eq:Sxilin}
S_{\xi} & = & 
   \int_0^\tau \mathrm{d}t \, \left[ C -2 U( \xi) \right]
\nonumber\\
 & = & \int_0^\tau \mathrm{d}t \, \left[ \frac{4\gamma_0}{\vartheta\tau^2}\bigl(x_\tau+X\bigr) + \chi\frac{\vartheta^2/\gamma_0}{T_0+\vartheta \xi} \right]\, ,
\end{eqnarray}
where in the second step we have used 
the explicit form (\ref{eq:T}) of the linear temperature profile.
The first term of the integrand of the cost is constant whereas the second one
depends on the position $\xi$. In fact, the instantaneous cost is lower at
 high temperatures which, in this case, are reached for large values of $\xi$.
When the transport time is fixed and long it can therefore be profitable to
spend part of it where the anomalous term is less costly even though
this is away from the target.

\begin{figure}
   \includegraphics[width=8.0cm]{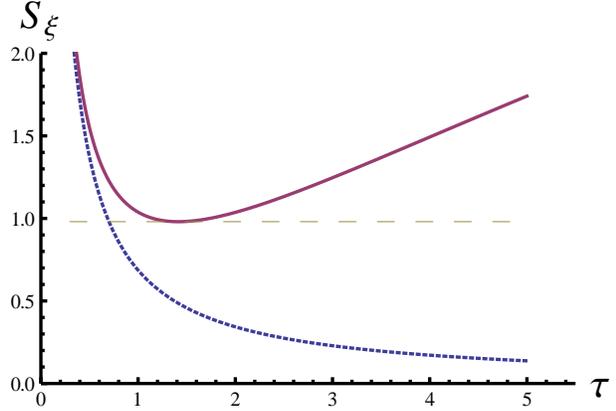}
\caption{Optimal cost between position $x_0=0$ and and $x_\tau=1$ as a function of process duration $\tau$
for $T_0=1$, $\vartheta=1$, $\gamma_0=1$. 
The dotted line refers to the case without the entropic anomaly
(solved by geodesics) and the solid line to the anomalous case of
inhomogeneous temperature. The reference dashed line indicates the
minimum cost for the anomalous case achieved for a time $\tau=\frac{\sqrt{2}(x_\tau-x_0)\gamma_0}{\vartheta}=\sqrt{2}$.}
 \label{fig:tau}
\end{figure}

It is now interesting to consider the dependence of the total cost
on the duration of the transport operation $\tau$ (Fig. \ref{fig:tau}). 
Naively we would expect a slow, quasi-static process (long $\tau$)
to be less dissipative and therefore associated with a lower
entropy production. For the non-anomalous setting of constant
temperature this is indeed the case (see \cite{aurell12a}) as we have 
\begin{equation}
 \label{eq:optimalreg}
 S_{\xi}(\tau)=\frac{4\gamma_0}{\vartheta \tau}\bigl(x_\tau+X|_{\chi=0}\bigr)\sim \frac{1}{\tau}\,.
\end{equation}
When a temperature gradient is present the situation changes drastically
as the anomalous contribution increases with the process duration and the minimum cost is
achieved at finite time. For the discussed linear temperature profile the optimal cost reads
\begin{eqnarray}
 \label{eq:optimalanom}
 S_{\xi}(\tau)&=&\frac{4\gamma_0}{\vartheta \tau}\left(x_\tau+X|_{\chi=1}\right)\nonumber\\ 
                &+& \sqrt{2}\log{\left[\frac{1-\left(\frac{x_\tau+X|_{\chi=1}-\frac{\vartheta\tau}{\gamma_0\sqrt{2}}}{x_\tau-x_0} \right)^2}
{1-\left(\frac{x_\tau+X|_{\chi=1}+\frac{\vartheta\tau}{\gamma_0\sqrt{2}}}{x_\tau-x_0} \right)^2}\right]}
\end{eqnarray}
where we recall that $X|_{\chi=1}$ also depends on $\tau$ as specified in \eqref{eq:X}.
This expression is not a monotonic decreasing function of $\tau$ and therefore, there is a finite valued
$\tau$ minimizing it: 
\begin{equation}
 \tau^*=\frac{\sqrt{2}(x_\tau-x_0)\gamma_0}{\vartheta}\,.
\end{equation}
It is interesting to note that this optimal duration of the protocol depends inversely on the intensity of the gradient
and corresponds to the case in which the solution of \eqref{eq:xi} is a
straight line. Furthermore, considering \eqref{eq:overshhot}, we can see that the overshooting
of the final target takes places for transport times that are longer 
than the optimal ones $\tau_{\mathrm{o.s.}}=2\sqrt{\frac{x_\tau+T_0/\vartheta}{(x_\tau-x_0)}}\tau^*>2\tau^*$.
Before moving to the conclusion we wish to recall that, although the solutions \eqref{eq:xi} are sufficient to assess the optimal transport duration and 
highlight several peculiarities of the optimal protocol in presence of temperature gradients, they still depend on 
the explicit evolution of the probability density (via the current velocity). In order to have a complete
solution for the protocols one has to consider the specific initial distribution and solve the
corresponding assignment problem. 

\section{Conclusion}
We have shown that the optimization of entropy production for driven diffusion processes
in an inhomogeneous temperature environment can be mapped into an auxiliary
deterministic transport problem describing motion on a curved manifold.
The metric tensor of the manifold is given by the inverse of the diffusion matrix.
Contributions to the entropy production due to the ``entropic anomaly''
\cite{celani12} play the role
of a potential energy for the auxiliary deterministic dynamics on
the curved manifold. In non-anomalous cases the optimization reduces to the solution of geodesics.
Recently, geodesics were found as optimal solutions
in control parameter space 
for excess power in \cite{sivak12} within a linear response analysis,
and for slowly varying protocols driving a particle in
a harmonic potential in \cite{zulkowski12}.
Using the simple example of one-dimensional diffusion
in a linear temperature gradient, we demonstrated that optimization of entropy
production including the anomaly requires finite processing times which
are inversely proportional to the gradient.
In contrast, for regular settings (homogeneous temperature), the optimal average entropy production
is reached in the quasi-static limit of adiabatically slow operation.
We have also shown that for slow transports the 
anomalous optimal trajectory is markedly different from the regular one
and may display non trivial features such as overshooting of the final target.
We have here presented the details in the case of a space-dependent temperature and
friction coefficient. However,
temperature and viscosity (friction) variations with time can be
treated along similar lines, the main difference to our central result
(\ref{eq:EL}) being an 
additional term from the time-derivative of the metric tensor.

\acknowledgements
This work was supported by the Academy of Finland as part of its
Finland Distinguished Professor program, project 129024/Aurell,
and through the Center of Excellence COIN. We furthermore acknowledge
financial support from the VR grant 621-2012-2982.

\end{document}